# Spectroscopic measurement of the refractive index of ion-implanted diamond

Alfio Battiato,[1,2] Federico Bosia,[1,2*] Simone Ferrari,[1] Paolo Olivero,[1,2] Anna Sytchkova,[3] Ettore Vittone[1,2]

[1] *Department of Experimental Physics- NIS Centre of Excellence, Università di Torino, Italy*
[2] *INFN Sezione di Torino and CNISM, Italy*
[3] *ENEA, Optical Coatings Group, via Anguillarese 301, 00123 Roma, Italy*
*\*Corresponding author: fbosia@to.infn.it*

We present the results of variable-angle spectroscopic ellipsometry and transmittance measurements to determine the variation of the complex refractive index of ion implanted single-crystal diamond. An increase is found in both real and imaginary parts at increasing damage densities. The index depth variation is determined in the whole wavelength range between 250 and 1690 nm. The dependence from the vacancy density is evaluated, highlighting a deviation from linearity in the high-damage-density regime. A considerable increase (up to 5%) in the real part of the index is observed, attributed to an increase in polarizability, thus offering new microfabrication possibilities for waveguides and other photonic structures in diamond.

Single-crystal diamond has attracted considerable attention in recent years in the photonics community due to the properties of its broad range of active luminescent centers, which offer promising opportunities in quantum cryptography and quantum information processing [1]. Nitrogen-Vacancy (NV) defects in particular display particularly attractive characteristics due to their individual addressability, optical spin polarization and long coherence times, even at room temperatures [2, 3]. The prospect of creating all-diamond integrated photonic structures has thus triggered interest in the possibility of fabricating optical waveguides [4] and other photonic structures [5] in diamond using ion implantation to modulate its refractive index. Moreover, an accurate control of refractive index variations is mandatory in all photonic applications which are based on ion implantation [6]. Despite this, a surprisingly small number of publications in the literature have addressed the problem of the index of refraction variation in diamond with ion irradiation [7-9]. Following the first systematic studies of damage-induced refractive index variation at a fixed wavelength [10] and the demonstration of waveguide fabrication in diamond with ion implantation [4], the dependence of refractive index variation upon structural damage needs to be systematically explored in a wide wavelength range for a broader spectrum of ion species and energies. In this paper, we report on the use of Variable-Angle Spectroscopic Ellipsometry [11] (VASE) integrated with optical absorption measurement to assess refractive index and extinction coefficient variations as a function of damage density.

**Experimental.** Ion implantation was performed on single-crystal CVD diamonds produced by ElementSix. The samples have 100 crystal orientation, size 3×3×0.5 mm$^3$ and are classified as type IIa, with two optically polished opposite large faces. Four samples were implanted with 180 keV B ions, at the Olivetti I-Jet facilities in Arnad (Aosta, Italy). The whole upper surface of the four samples was irradiated uniformly with fluences of $10^{13}$, $5·10^{13}$, $10^{14}$ and $5·10^{14}$ cm$^{-2}$ with an accuracy below 0.5%. Ellipsometric characterization of the samples was performed using a Woollam M2000-FI [12] variable-angle spectroscopic ellipsometer in the wavelength range from 246 to 1690 nm. For each sample, data were acquired at incidence angles of 63°, 65°, 67°, 69° and 71° with respect to the surface normal.

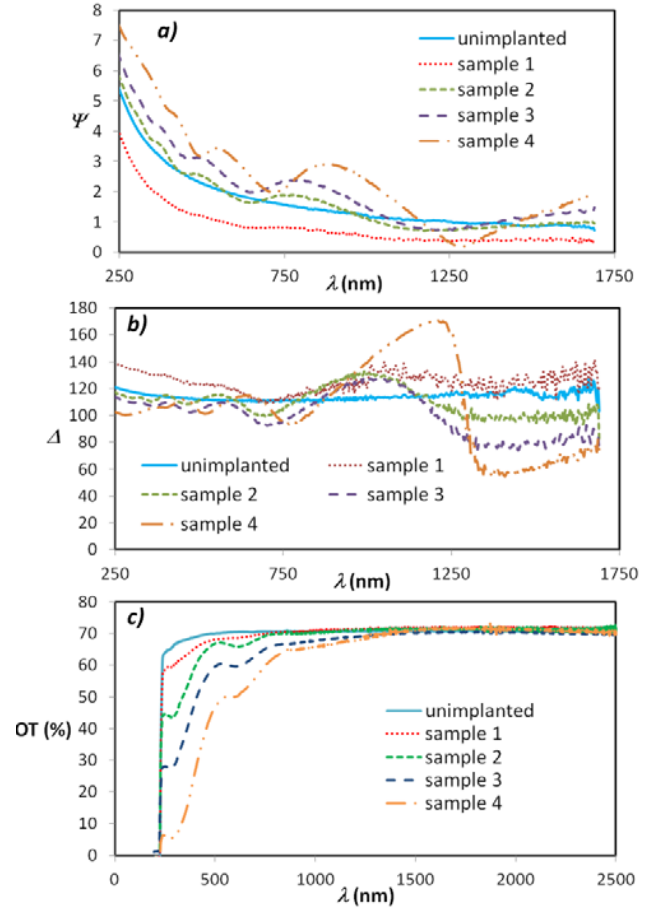

Fig. 1. (Color online) a) Amplitude ratio $\Psi$ and b) phase $\Delta$ spectra for unimplanted and implanted samples (1-4, relative to fluences of $1·10^{13}$, $5·10^{13}$, $1·10^{14}$, $5·10^{14}$ ions cm$^{-2}$, respectively). Only results for an incident angle of 67° are shown for clarity. c) corresponding optical transmission (OT) spectra.

An example of spectra of the measured ellipsometric angles Psi ($\Psi$) and Delta ($\Delta$) at an impinging angle close to the Brewster angle of 67° is shown in Figs.1a and 1b for the 4 implanted samples and compared to a reference spectrum for an unimplanted diamond sample. The detected oscillations increase with fluence, and are typical of the interference arising in thick transparent films on a substrate [11]. The Optical Transmission (OT) measurements were performed at normal incidence with a Perkin Elmer Labmbda 950 spectrophotometer in the range 250-3300 nm, with a spectral resolution of 1 nm (Fig. 1c). A suitably developed spatial masking system was applied for a precise determination of the illuminated zone, ensuring a light spot size of approximately 1 mm by 1.5 mm. The spectrophotometer operates in a double-beam mode with a photometric accuracy better than 0.00015 (0.015 %OT).

**Analysis.** The VASE data were fitted using Complete EASE v.4.17 software from Woollam Inc. [12] to derive the depth variation of the index of refraction n. Having ascertained the small influence of absorption (see discussion below), a multilayer model for the damaged substrate was adopted with a Cauchy-type spectral dependence for $n(\lambda)$ with fixed layer thicknesses chosen between 30 and 50 nm, top surface roughness considered as a mixture of 50%-50% volume fraction for air [13] and diamond, and backside reflections included in the fit.

The obtained n values in the layers were thus compared to the vacancy density depth profiles, calculated by means of Monte Carlo simulations using the TRIM [14] code (with an atom displacement energy for diamond of 50eV). To account for damage accumulation and saturation effects, the Crystal-TRIM (C-TRIM) code [15] was also employed. As an example, results for n vs. depth z in sample 3 at $\lambda$ =638 nm are shown in Fig. 2. The consistency between experimental data and C-TRIM numerical results is very satisfactory, particularly considering that in the analysis of the ellipsometric data no preliminary assumptions were made about the variation of n. On the other hand, a discrepancy with TRIM results is highlighted in Fig. 2. Similar trends are obtained for all samples and wavelengths in the considered range.

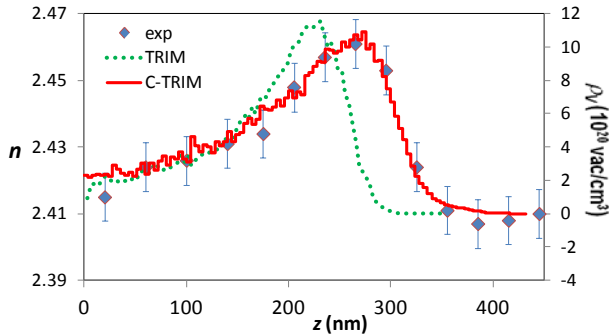

Fig. 2. (Color online) Measured refractive index n depth variation compared to TRIM- and Crystal-TRIM-calculated vacancy density profiles for sample 3 ($\phi$= $10^{14}$ions cm$^{-2}$) at 638 nm.

The determined n values for all layers of 3 of the 4 implanted samples (the lowest fluence yields a negligible index variation) are reported in Fig. 3 as a function of the corresponding vacancy density values for wavelengths of interest for diamond, e.g. for $\lambda$=638 nm, corresponding to the zero-phonon line of NV$^-$ centers, and $\lambda$=1550 nm, corresponding to the "third" optical window in telecommunications. A deviation from linearity at high damage densities is apparent. In the same graph we also report the linear relationship between n and the vacancy density $\rho_V$ as evaluated by means of interferometric techniques [10] for the same wavelength, which was found to be valid for $\rho_V$< 3·$10^{21}$ cm$^{-3}$. It is apparent that in the present case the linear relationship adequately describes the behavior only for a density of vacancies $\rho_V$< 2·$10^{21}$ cm$^{-3}$, whilst for 3·$10^{21}$ cm$^{-3}$< $\rho_V$< $10^{22}$ cm$^{-3}$ the trend is sub-linear, and the data can be interpolated by an empirical exponential-like curve, e.g.: $n = n_0 + n_\infty [1 - \exp(-\rho_V/b)]$, where $n_0$ is the refractive index of the undamaged diamond, and $b$ and $c$ are fitting constants. The latter curve is also reported in Fig. 3 with $c$=0.1 and $b$=1.5·$10^{21}$ cm$^{-3}$ for $\lambda$=638 nm and $c$=0.09 and $b$=1.5·$10^{21}$ cm$^{-3}$ for $\lambda$=1550 nm. The slight discrepancy between these results and those reported in previous works [10] can be ascribed to the different stress state in the damaged diamond: here, the whole sample surface is uniformly implanted and is therefore free to expand as amorphization due to ion implantation occurs, while the same cannot be said for the previously considered 125×125μm$^2$ implanted areas [10], where the constrained expansion causes a stress build up and possibly a greater increase in index variation. Moreover, it is worth stressing that significantly different ion species and energies have been employed here. Having associated refractive index to vacancy density values, it is possible to analyze the spectral behavior of the data for given damage densities. Results for n are shown in Fig. 4a for $\rho_V$=2·$10^{20}$ cm$^{-3}$, 13·$10^{20}$ cm$^{-3}$, and 50·$10^{20}$ cm$^{-3}$.

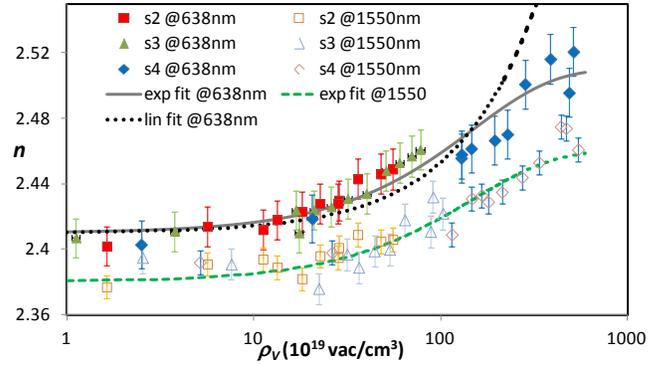

Fig. 3. (Color online) Overall refractive index ($n$) values for samples 2 to 4 ("s2", "s3", "s4") at 638 nm and 1550 nm as a function of calculated vacancy density $\rho_V$ (log scale): the dotted line is the linear (lin) fit of data relevant to 638 nm; the continuous ad dashed lines are exponential (exp) fits at 638 and 1550 nm, respectively

The imaginary part of the refractive index $k$ was evaluated through the analysis of the transmission spectra in normal geometry. Here, the thickness variation of $k$ is not analyzed, and an "equivalent" single layer with constant absorption depth profile is considered." The absorption coefficient $\alpha_\phi(\lambda)$ of the implanted layer of thickness $d$ for a given implantation fluence $\phi$, can be estimated as:

$$\alpha_\Phi(\lambda) - \alpha_0 = -\frac{1}{d} \cdot \ln\left[ \frac{1-\Gamma^2}{2\cdot t \cdot \Gamma^2} \cdot \left( \sqrt{1 + \frac{4\cdot t^2 \cdot \Gamma^2}{(1-\Gamma^2)^2}} - 1 \right) \right] \quad (1)$$

where

$$\alpha_0 = -\frac{1}{D} \ln\left( \frac{(1-R_0)^2}{2T_0 R_0^2} \cdot \sqrt{1 + \frac{4T_0^2 R_0^2}{(1-R_0)^4}} - 1 \right) \quad (2)$$

is the absorption coefficient [9, 16], $T_0$ the transmittance and $R_0$ the reflectance of the pristine (subscript 0) sample,

$\Gamma = R_0 \cdot \exp[-\alpha_0 \cdot D]$, $t = T_I/T_0$, $T_I$ the transmittance of the irradiated sample. The sample thickness $D$ is 500 μm and the thickness $d$ of the region optically modified by irradiation is around 400 nm, as shown in Fig. 2.

The average extinction coefficient $k_\phi(\lambda)$ spectra, evaluated through the expression $k_\phi(\lambda) = \lambda \cdot \alpha_\phi(\lambda)/4\pi$, are shown in Fig. 4b. The spectra are dominated by bands located at about 2 eV and 4.2 eV. The former can be attributed to vacancy-related GR1 centers, as observed by other authors in irradiated IIa diamonds and in amorphous carbon films [17], [18]. The broad band at 4.2 eV can be deconvoluted in three lorentzians centered at 3.5, 4.2, 5.2 eV and has been tentatively attributed to vacancy-related GR2-8 centers in previous works [19]. Finally, it is worth noting that in the main absorption bands, $k_\Phi$ is limited to a few percent of the refractive index; thus justifying the adoption of the Cauchy model to evaluate the real part of the refractive index ($n$) from the ellipsometric measurements. Indeed, inclusion of the obtained $k$ in ellipsometric fits does not change the result for $n$ significantly. Thus, the retrieval of real and imaginary parts of the complex refractive index may be considered almost separated tasks.

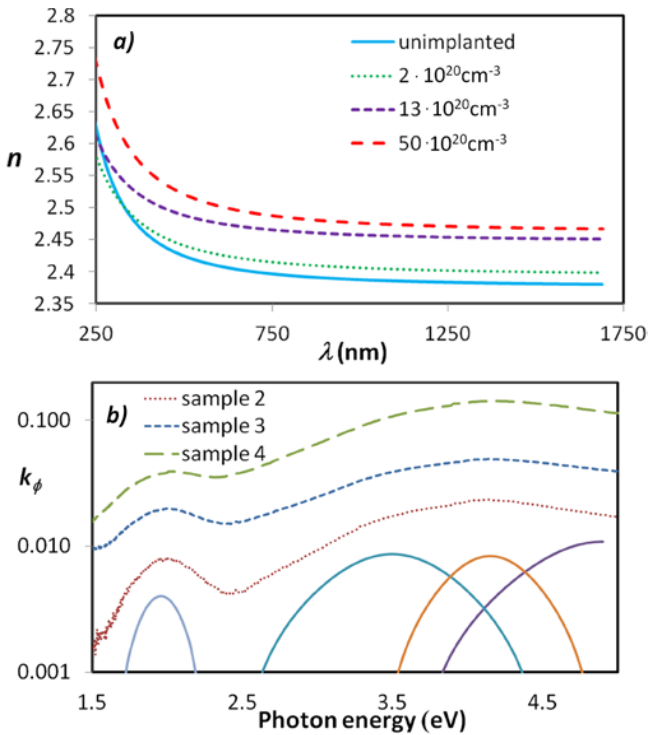

Fig. 4. (Color online) (a) Spectral variation of the measured refractive index for different vacancy densities. (b) Extinction coefficient k spectra from different irradiation fluences. The continuous lines are the lorentzian components of the spectrum at the lowest fluence.

**Conclusions.** In conclusion, we presented a systematic investigation on the variation of the refractive index of ion-implanted single crystal diamond over a broad spectral range. We observed an increasing trend of $n$ at increasing damage densities, with a deviation from linearity in the high-damage range. An increase in the refractive index of ion-implanted birefringent crystals has been found to occur in some cases (e.g. LiNbO$_3$) for the extraordinary index, with a concurrent decrease of the ordinary index [18]. In the case of non-birefringent crystals, to the best of the authors' knowledge, such an index increase has only been found in Nd:YAG [20, 21]. Given the considerable mass density decrease of diamond after ion implantation due to induced structural damage, and having excluded stress as the main cause for the index increase (see above), the effect can only be ascribed to an increase in polarizability [18]. Further experimental studies and ab initio simulations may help clarify this hypothesis and address issues like stress dependence, induced birefringence, etc., with the aim of exploiting results for the fabrication of efficient all-diamond optical structures.

**Acknowledgments.** The authors wish to thank Dr. P.Schina for specimen preparation at the Olivetti I-JET facilities at Arnad (Aosta). P. O. is supported by the "Accademia Nazionale dei Lincei – Compagnia di San Paolo Nanotechnology grant".